\documentclass[paper=A4,fontsize=12pt]{scrartcl}
\usepackage[dvipdfm] {graphicx}
\usepackage{amssymb,amsmath}
\usepackage{siunitx}
\begin{document}

\begin{center}
{\Large
KamLAND-PICO PROJECT TO SEARCH \\FOR COSMIC DARK MATTER
}

\vspace{0.5cm}
K.Fushimi$^{1}$, Y.Awatani$^{1}$, H.Ejiri$^{2}$, R.Hazama$^{3}$, H.Ikeda$^{4}$, 
K.Imagawa$^{5}$, K.Inoue$^{4}$, A.Kozlov$^{6}$, R.Orito$^{1}$, T.Shima$^{2}$, R.Sugawara$^{1}$ 
and K.Yasuda$^{5}$ \\

\vspace{0.5cm}
1) Institute of Socio, Arts and Sciences, The University of Tokushima, 1-1 Minamijosanjimacho
 Tokushima city, Tokushima, 770-8502, JAPAN

2) Research Center for Nuclear Physics, Osaka University, 10-1 Mihogaoka Ibaraki city, 
Osaka, 567-0042, JAPAN

3) Graduate School and Faculty of Human Environment, Osaka Sangyo University, 
3-1-1 Nakagaito, Daito city, Osaka, 574-8530, Japan 

4) Research Center for Neutrino Science, Tohoku University 6-3, Aramaki Aza Aoba,Aobaku Sendai city, 
Miyagi,980-8578,JAPAN 

5)  I.S.C. Lab., 7-7-20 Saito Asagi Ibaraki city, 567-0085 Osaka, JAPAN

6) Kavli Institute for the Physics and Mathematics of the Universe,
5-1-5 Kashiwanoha, Kashiwa city, 277-8583, Japan

\end{center}

\begin{abstract}
KamLAND-PICO project aims to search for WIMPs dark matter by means of NaI(Tl) scintillator.
To investigate the WIMPs candidate whose cross section is as small as $10^{-9}$ pb, 
a pure NaI(Tl) crystal was developed by chemical processing and taking care of surroundings.
The concentration of U and Th chain was reduced to $5.4\pm0.9$ ppt and $3.3\pm2.2$ ppt, respectively.
It should be remarked that the concentration of $^{210}$Pb which was difficult to reduce reached to 
the high purity as $58\pm26$ $\mu$Bq/kg.
\end{abstract}
\section{Introduction}
\subsection{Brief introduction of cosmic dark matter}
\begin{figure}[ht]
\centering
\includegraphics[viewport=0 0 494 284,width=0.8\textwidth]{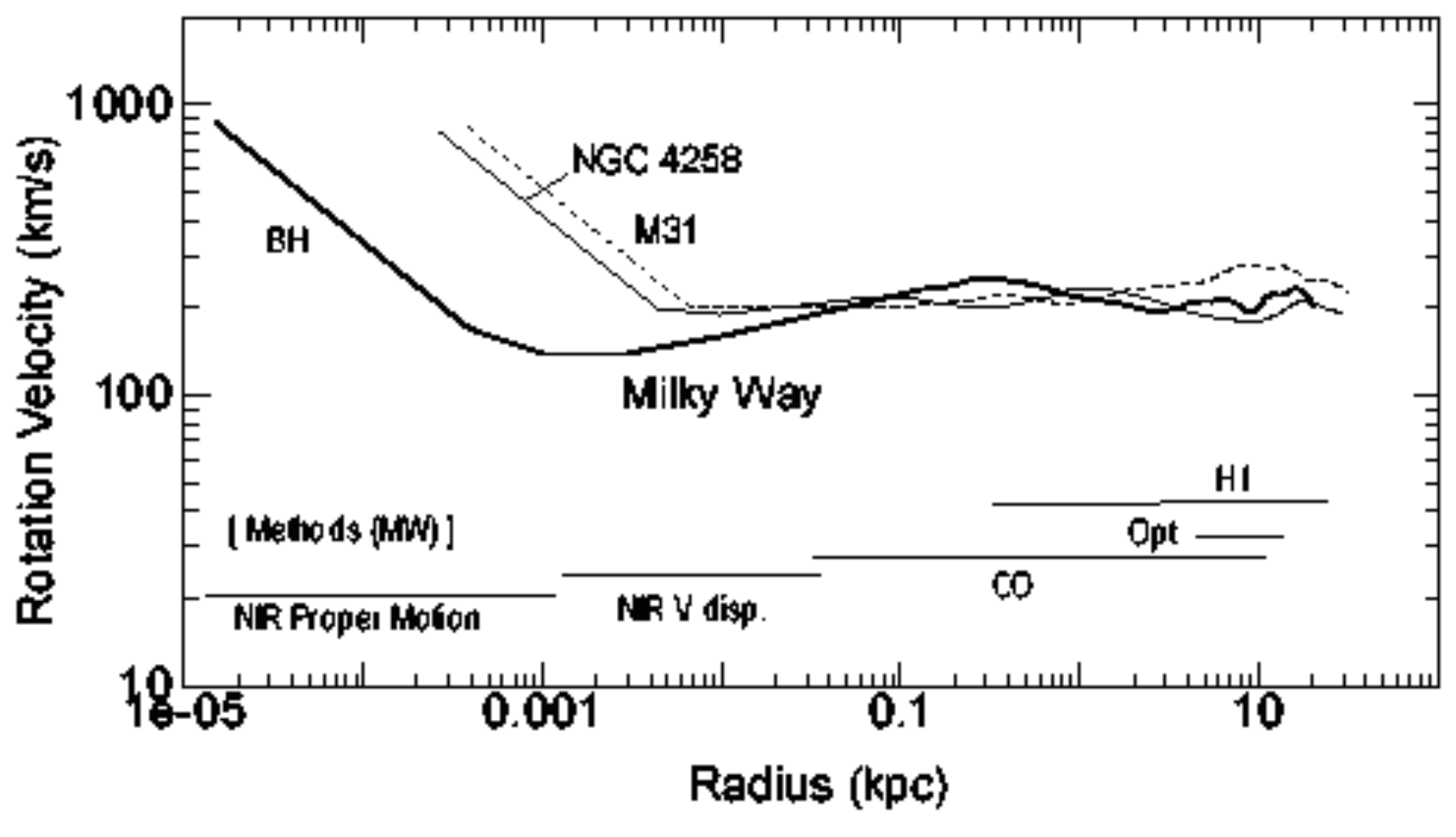}
\caption{Rotation curve for several galaxies. The solid line is the rotation curve of our galaxy\cite{Sofue}.}
\label{fg:rotation}
\end{figure}

The cosmic dark matter (DM) is an unknown matter which does not emit nor absorb any photon.
There are many evidences for DM by various cosmological observations.
The first evidence for DM was addressed by observing the motion of galactic clusters\cite{Zwicky}.
The existence of the invisible mass in our galaxy was also reported \cite{Sofue}.
The ratio of the kinetic mass of the galaxy to the luminous mass of the galaxy gives an evidence of invisible matter. 
The luminous mass is determined by the total luminosity by stars and gas in the galaxy.
The kinetic mass is determined by observing the rotation speed of the galaxy.
The distribution of rotation speed in galaxies are shown in Fig.\ref{fg:rotation}.
The large amount of invisible mass distribution far from the galactic center was concluded by 
the flat behavior of the rotation speed.

The indirect evidence for invisible mass was given by the observation of the cosmic microwave background (CMB).
 CMB is the afterglow of the Big Bang which were emitted after 400,000 years after the beginning of the Universe.
The initial temperature of the CMB corresponds to the recombination epoch of the Universe, about 
3000 K.\@
The wavelength of the CMB was stretched by the expansion of the Universe to the present temperature
which corresponds to 2.726 K.

The small deviation of the wavelength corresponds to the fluctuation of the density of matter in the 
region of the Universe. 
The dense regions in the early Universe evolved to the structures that we see today;
galaxies, clusters and superclusters.
The angular size of the fluctuation gives the important information on the structure of the Universe
because the size depends on the curvature of the space between the emission point and the 
observer.

The Planck observed the CMB structure with the high angular resolution \cite{Planck}.
The Planck data determine the cosmological parameters to high precision from the CMB.\@
The density parameter $\Omega$ of the Universe and the Hubble constant $H_{0}$ are
listed below.
\begin{align}
\Omega_{b}h^{2} &= 0.02205\pm 0.00028 \\
\Omega_{c}h^{2} &= 0.1199 \pm 0.0027 \\
\Omega_{\Lambda} & =0.686 \pm 0.020 \\
H_{0} & = (67.3 \pm 1.2) \rm{\ km s^{-1} Mpc^{-1}}
\end{align}
Where $\Omega_{b}$, $\Omega_{c}$ and $\Omega_{\Lambda}$ are density parameters of 
baryon, cold dark matter (CDM) and cosmological constant, respectively.
$h\equiv H_{0}/(100$ km/sec/Mpc) is the Hubble parameter.
The composition of the Universe cannot be dominated by baryons that we usually observe in the Universe.
Other component are unknown dark matter and unknown dark energy.

\subsection{Candidates for cosmic dark matter and their signal}
The dark matter is supposed to be unknown elementary particles that are proposed by the theories beyond 
the standard theory of elementary particles.
From the view point of cosmology, the main component of dark matter is supposed to be a non-relativistic 
particle which was decoupled from a cosmic radiation after the Universe became cold.
These particle is called the cold dark matter (CDM) and plays an important role in creating a large scale 
structure of the Universe.
There is, on the other hand, hot dark matter (HDM) which was decoupled from 
a cosmic radiation during the Universe was hot.
Since the HDM particles travels with relativistic speed, they destroy the large scale structure.
Consequently, the HDM cannot be a main component of the dark matter.
The candidates for CDM particles are axion and WIMPs (Weakly Interacting Massive Particles).
The KamLAND-PICO project is suitable for the detection of both axion and WIMPs.
In the present paper, we focus on the detection of WIMPs. 

WIMPs are the weakly interacting massive particles whose mass is expected to between 10 GeV$/c^{2}$ 
and 1000 GeV$/c^{2}$ where $c$ is the speed of light.
The cross section of WIMPs-nucleus interaction is classified into three types, spin-independent (SI), 
spin-dependent (SD) and nuclear excitation (EX)\cite{Fushimi_PL}.
The cross section dependence are briefly given as,
\begin{align}
\sigma_{SI} & \propto A^{2} \\
\sigma_{SD} & \propto C^{2}J(J+1) \\
\sigma_{EX} & \propto \frac{J+1}{J'+1}\left|\frac{p_{f}}{p_{i}}\right| .
\end{align}
Where $A$ is mass number of the target nucleus. $J$ and $J'$ are spins of the 
ground state and excited state of the target nucleus, respectively.
$p_{i}$ and $p_{f}$ are the initial and the final momenta of incident WIMP.

The mass density $\rho_{D}$ and the mean velocity $v_{D}$ of the WIMPs in the vicinity of the solar system 
has been calculated by observing the stellar motion of the Galaxy\cite{MNRAS_Smith}.
The obtained value of $\rho_{D}$ by various observation ranges from 0.3 to 0.5 GeV$/c^{2}/$cm$^{3}$.
The mean velocities range from 200 to 230 km/sec.
In the present paper, the conservative values $\rho_{D}=0.3$ GeV$/c^{2}$/cm$^{2}$ and 
$v_{D}=230$ km/sec are used for the analysis of WIMPs sensitivity. 

The signal for WIMPs in a radiation detector is given by a nuclear recoil by elastic or inelastic scattering.
The energy spectrum of recoil nucleus observed in a detector is given by \cite{Lewin}
\begin{equation}
\frac{dR}{dE_{ee}}=\frac{1}{f}\frac{R_{0}}{rE_{0}}\exp\left(-\frac{E_{R}}{rE_{0}}\right)\left|F(q)\right|^{2},
\end{equation}
where, $R_{0}$ is the total event rate of WIMPs-nucleus elastic scattering, 
$f$ is the quenhing factor of detector response and $\left|F(q)\right|^{2}$ is the form factor.
$E_{0}=m_{D}v_{D}^{2}/2$ is the mean kinetic energy of WIMPs and $m_{D}$ is WIMPs mass.
$r=4m_{D}m_{N}/(m_{D}+m_{N})^{2}$ and $m_{N}$ is the mass of target nucleus.

The electron equivalent energy $E_{ee}$ is the observed energy which is calibrated by electron scattering.
The nuclear recoil energy $E_{R}$ is the kinetic energy of recoil nucleus.
The quenching factor for nuclear recoil $f$ is the important factor to estimate the energy spectrum of WIMPs.
The detector response in a scintillator and semiconductor by nuclear recoil is much smaller than the 
one by electron with the same kinetic energy.
A nuclear recoil gives only a small fraction of its kinetic energy to scintillation and ionization processes.
The quenching factor $f$ is determined by the ratio of the scintillation output by nuclear recoil ($N_{nr}$) 
to the one by electron ($N_{e}$) with the same kinetic energy\cite{Lindhard} as 
\begin{equation}
f\equiv \frac{N_{nr}}{N_{e}}=\frac{E_{ee}}{E_{R}}.
\end{equation}
The values for various detectors were measured by neutron scattering\cite{Fushimi_PRC, BPRS,DAMA_Xe}.
The energy threshold of the detector must be as low as possible to observe the WIMPs signal efficiently.

\section{ Present status of WIMPs search in the world}
Many groups in the world are applying various types detector to search for WIMPs.
The major experimental groups in the world are listed in table \ref{tb:Groups}.
\begin{table}[ht]
\caption{Dark matter searchers in the world. There are many other groups in the world.}
\label{tb:Groups}
\centering
\begin{tabular}{l|lllc}\hline
Group &  Country & Target & Status & Reference \\ \hline
LUX &USA & Xe &  No signal. Stringent limit.& \cite{LUX_1st} \\
Xenon100 & USA & Xe & No signal. Stringent limit.& \cite{Xenon100} \\
CoGeNT & USA & Ge & Light WIMPs signal. & \cite{CoGeNT2011} \\
CDMS & USA & Ge, Si & Three candidate events. & \cite{CDMS2013} \\
CRESST & LNGS & CaWO$_{4}$ & Light WIMPs signal. & \cite{CRESST} \\
DAMA/LIBRA & LNGS & NaI(Tl) & 13 annual mudolation signal. & \cite{DAMA2013}\\
EDELWEISS & Modane & Ge & Three candidate events. & \cite{EDELWEISS-II} \\
ANaIs & Canfranc & NaI(Tl) & Suffered by $^{210}$Pb contamination. & \cite{ANaIS2010} \\
DM-ICE & South Pole & NaI(Tl) & Construction. & \cite{DM-ICE} \\ 
KIMS & Korea & CsI(Tl) & No signal. Stringent limit.& \cite{KIMS}\\
XMASS & Kamioka & Xe & No signal. Stringent limit.& \cite{XMASS_LightWIMPs} \\
NewAge & Kamioka & CF$_{4}$ & R\&D & \cite{NewAge} \\
ANKOK & Kamioka & Ar & R\&D& \cite{ANKOK} \\
KamLAND-PICO & Kamioka & NaI(Tl) & R\&D & This work. \\
\hline
\end{tabular}
\end{table}
CDMS, CoGeNT and EDELWESS reported the signal for light WIMPs whose mass was around 10 GeV$/c^{2}$.
The oldest report on the candidate signal for WIMPs was given by DAMA/LIBRA\cite{DAMA_First}.
They continuously observed an annual modulating signal for 13 annual cycles with more than $8\sigma$ 
significance. 
They excluded another candidates causing the modulating signal such as fluctuation of background and 
electronics.
However, the allowed regions of the mass and cross section of WIMPs do not agree with each other 
and all the region of expected WIMPs signal has been ruled out by the experiments by using Xe.
The situation of WIMPs search is now confusing.

Many groups have been applying a NaI(Tl) scintillator to search for dark matter.
However, the sensitivities given by the groups does not compare to the one given by DAMA/LIBRA 
because their sensitivities are suffering from the impurity of NaI(Tl) crystal.
The densities of radioactive impurity in the NaI(Tl) applied by DAMA/LIBRA were 
0.7$\sim$10 ppt for U-chain, 0.5$\sim$0.7 ppt for Th-chain, 5$\sim$30 $\mu$Bq/kg for $^{210}$Pb
and 20 ppb for potassium\cite{DAMA_NIM}.
The competing purity is needed to verify the annual modulating signal reported by DAMA/LIBRA.

\section{Outline of KamLAND-PICO project}
KamLAND-PICO aims to search for cosmic dark matter by means of highly radiopure NaI(Tl) scintillator
which is installed into KamLAND.\@
KamLAND is the neutrino and anti-neutrino detector using the largest and highly radiopure 
liquid scintillator (LS).
The detector is placed in Kamioka underground observatory.
The 1000 ton of ultra pure LS is contained in a 13 m diameter spherical balloon
made of 135 $\mu$m thick nylon/EVOH (ethylene vinyl alcohol copolymer) composite film\cite{KamLAND}.
1325 modules of 17 inch diameter photomultiplier tubes (PMTs) and 554 module of 20 inch diameter PMTs 
are installed on 18 m diameter stainless steel vessel.
The space between the stainless vessel and the LS balloon is filled with an ultra pure buffer oil.

The KamLAND detector system is the ideal active shield for the WIMPs detector.
The mineral oil acts as a neutron moderator for fast neutron which gives fake event of WIMPs signal.
The central LS has a good sensitivity to fast neutron, furthermore, it has an enough volume to absorb all the 
$\gamma$ ray background.

The PICO-LON detector consists of NaI(Tl) crystal.
The NaI(Tl) scintillator has great advantages to search for WIMPs because of the reasons below.
i) $^{127}$I is sensitive to SI type WIMPs because it has a large mass number. 
\ ii) Both $^{23}$Na and $^{127}$I are sensitive to SD type WIMPs because they have finite nuclear spins. 
\ iii) $^{127}$I is sensitive to nuclear excitation by SD type WIMPs because it has a low lying 
excited state ($E_{X}=57.6$ keV). 
\ iv) Both $^{23}$Na and $^{127}$I are contained in NaI with 100\% abundance.
\ v) A large volume NaI(Tl) detector is developed with low cost.

The final version of PICO-LON detector consists about thirty thousand plates of thin NaI(Tl) crystal
whose dimension is $15\times15\times0.1$ cm$^{3}$ and total mass is 1 ton.
PICO-LON detector system has a good sensitivity to all the types of interaction.
Especially, the coincidence measurement of 57.6 keV $\gamma$ ray and nuclear recoil enlarges the 
signal-to-noise ratio.
The thickness of the NaI(Tl) has been optimized to perform coincidence detection of 57.6 keV $\gamma$ 
ray and nuclear recoil.
The successful development of thin and wide area NaI(Tl) scintillator was already reported by 
previous papers\cite{Fushimi_thinNaI, Harada}.

The development of an extremely high purity NaI(Tl) crystal is needed to push forward the 
KamLAND-PICO project.
The first phase of KamLAND-PICO aims to purify the NaI(Tl) crystal and develop a large volume 
NaI(Tl) scintillator system to find a WIMPs signal.

\section{Development of pure NaI(Tl) crystal}
\subsection{Required purity for WIMPs search}
The sensitivity to WIMPs is suffered from radioactive impurities contained in a NaI(Tl) crystal. 
The following radioactive impurities must be reduced to their required concentrations in order to ensure 
the sensitivity to WIMPs.
The concentration of U-chain and Th-chain impurities must be lower than 1 ppt because their progeny 
emit many $\beta$ rays and $\gamma$ rays. 
The potassium impurity must be lower than 20 ppb. 

\begin{figure}[ht]
\centering
\includegraphics[viewport=0 0 167.68357  131.475,width=0.4\textwidth]{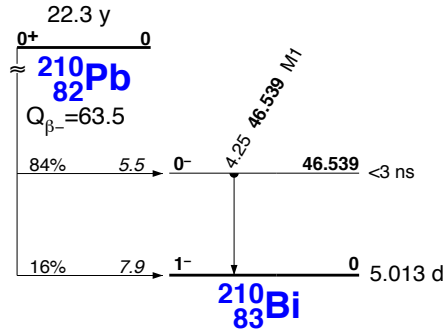}
\caption{A decay scheme of $^{210}$Pb.\cite{TOI}}
\label{fg:210Pb_scheme}
\end{figure}

The most important task to improve the sensitivity is to reduce the concentration of $^{210}$Pb and 
its concentration must be lower than 100 $\mu$Bq/kg.
A $\gamma$ ray whose energy is 46.5 keV and a $\beta$ ray whose maximum energy is 17 keV are 
simultaneously emitted (see Fig.\ref{fg:210Pb_scheme}).
Since these radiations makes a similar event to the inelastic excitation of $^{127}$I, 
the $^{210}$Pb reduces the sensitivity when the energy resolution of NaI(Tl) is worth than 30 \% at 60 keV
in full-width at half-maximum.
The progeny $^{210}$Bi also reduces the sensitivity by emitting a $\beta$ ray whose maximum energy is 1162 
keV and bremsstrahlung radiations.

\subsection{Purification of NaI(Tl) crystal}
It is difficult to reduce the concentration of $^{210}$Pb because its half life is four orders of magnitude 
shorter than that of $^{238}$U.
Because of the short half life, the number density of $^{210}$Pb ion corresponding to 100$\mu$Bq/kg is as small as
$1\times10^{5}$/kg.
The $^{210}$Pb was contaminated during drying process since the concentration of $^{210}$Pb was much larger 
than that of $^{226}$Ra in our previous NaI(Tl) crystal.
A significant amount of $^{210}$Pb is mixed into NaI(Tl) powder by dried air even if it contains a small 
density of $^{222}$Rn.

The $^{210}$Pb contained in a purchased NaI(Tl) powder was removed by using an ion-exchange resin.
The treated solution was dried by rotary evaporator.
The vacuum of the evaporator was broken by supplying pure nitrogen gas.
The air of the workshop was filtered by a HEPA filter 48 hours before the treatment and 
the filter was continuously operated.
The dried powder was filled in a highly pure crucible whose diameter was 15 cm.
The Bridgman method was applied to grow a NaI(Tl) crystal.
The crystal was ground and polished to 3 inch$\phi\times$3 inch cylinder and put into 
an aluminum housing.
One end of NaI(Tl) crystal was contacted with a Pyrex light guide and the other surfaces were covered 
with a PTFE reflector.

\subsection{Measurement of U and Th contamination}
The U chain and Th chain impurities were measured by applying the pulse shape discrimination (PSD) method
\cite{DAMA_NIM,Umehara}.
Measuring alpha rays ensures the precise determination of radioactivity in the NaI(Tl) crystal
because its range is short enough.
The alpha rays and beta/gamma rays were distinguished by the decay constant of the scintillation pulses.
The decay constant of NaI(Tl) of alpha ray event is 190 nsec and that of beta/gamma ray is 230 nsec.

\begin{figure}[ht]
\centering
\includegraphics[viewport=0 0 506.444 346.5921,width=0.4\textwidth]{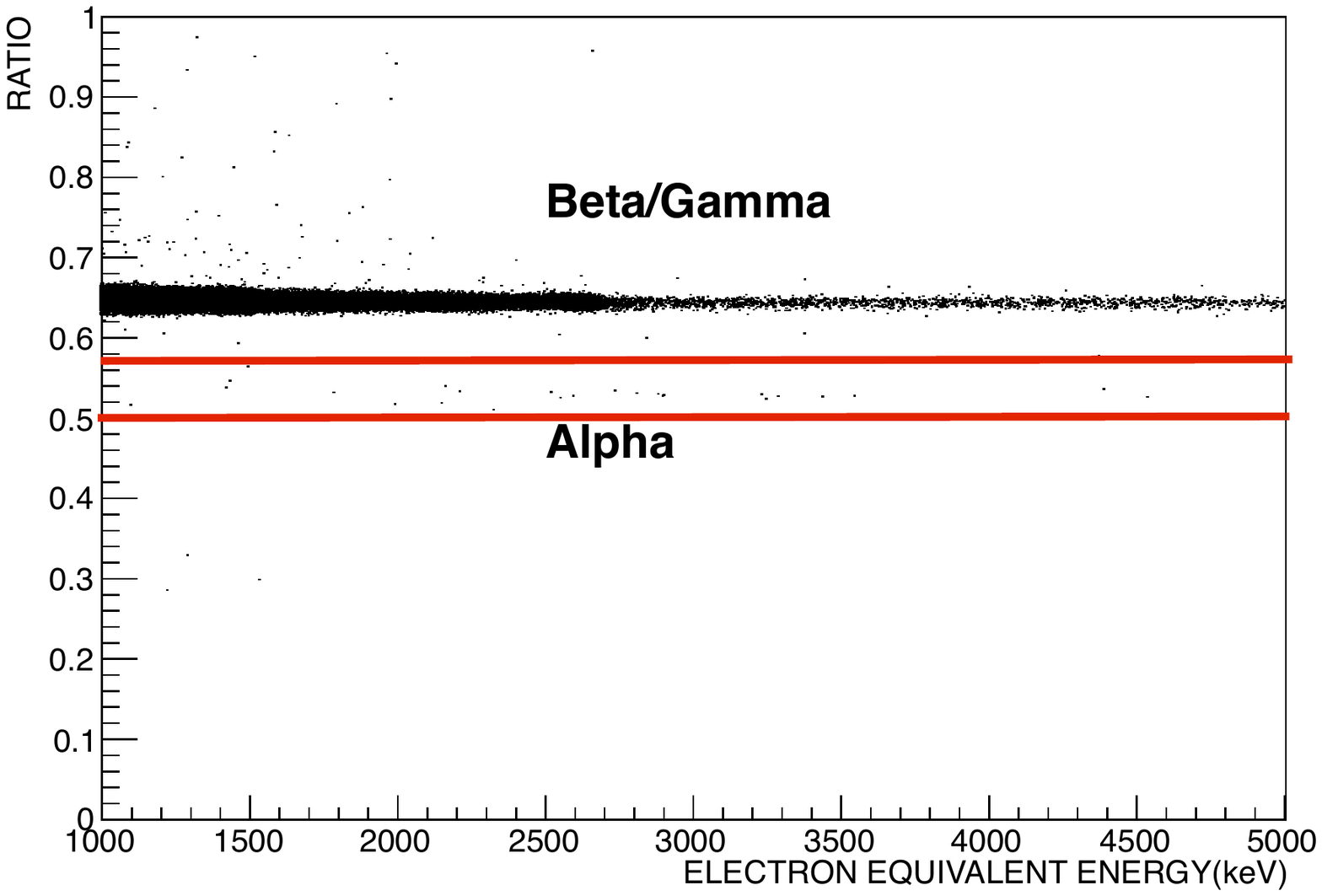}
\hspace{0.05\textwidth}
\includegraphics[viewport=0 0 496.1753 341.8402, width=0.4\textwidth]{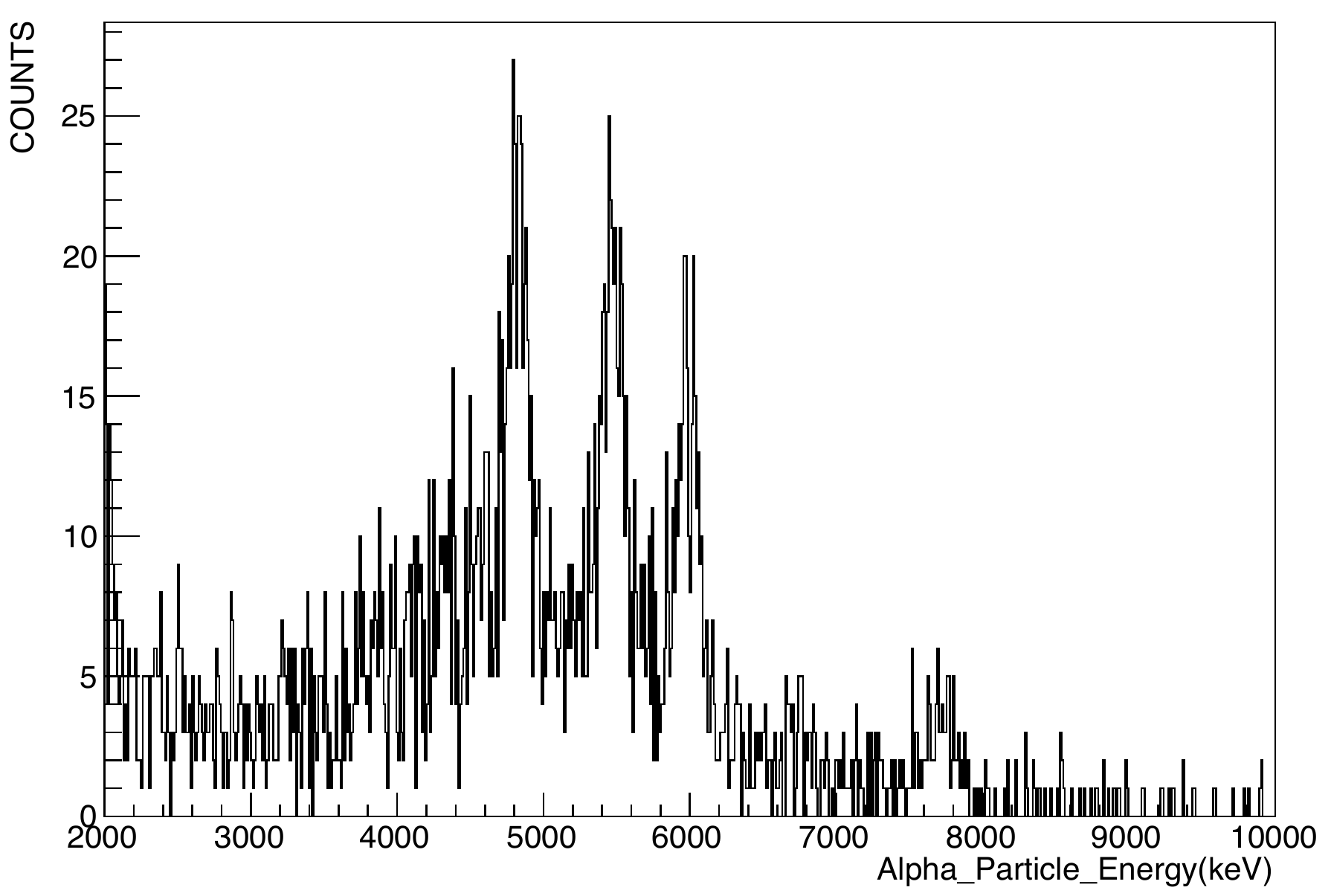}
\caption{Left: The conceptual drawing of PSD analysis performed by ingot 24. 
The horizontal axis is electron equivalent energy in NaI(Tl).
Right: The energy spectrum of alpha ray observed by ingot 23. 
}
\label{fg:PSD}
\end{figure}

A simple PSD method was constructed as described below.
A current output of photomultiplier tube (PMT) was divided into three routes by a simple signal divider.
The first signal was input into a discriminator to make a charge integrating 
analog-to-digital converter (CSADC) gate whose duration was 1 $\mu$sec.
The second one was input into the CSADC through 200 nsec cable delay.
The last one was input into a CSADC directly.
The timing of a gate signal for CSADC was optimized for the linear signal input through 
200 nsec cable delay.
The linear signal which was input directly into CSADC was integrated 200 nsec after the 
signal started.
The pulse shape difference was calculated by the ratio defined as 
\begin{equation}
Ratio \equiv \frac
{\int^{1200 ns}_{200 ns}I_{0}\exp\left(\frac{t}{\tau}\right)dt}
{D\int^{1000 ns}_{0 ns}I_{0}\exp\left(\frac{t}{\tau}\right)dt}.
\end{equation}
Where $I_{0}$ and $\tau$ are the initial current and the decay constant of signal, respectively.
The factor $D$ is an attenuation factor due to 200 nsec delay cable.
A conceptual drawing of extracting alpha ray events from a large number of background events is shown in 
Fig.\ref{fg:PSD}.
The events of beta/gamma ray lie in the range of $Ratio$ between 0.6 and 0.7 in Fig.\ref{fg:PSD}.
The alpha ray events lie between 0.5 and 0.6.
A small number of alpha ray events were clearly extracted.

In the right panel of Fig.\ref{fg:PSD}, the energy spectrum of the extracted alpha ray events is shown.
The data was accumulated for the live time of 26.7 days$\times$1.25 kg by ingot 23.
The densities of contamination in ingots 23 and 24 were the same as each other. 
Three prominent peaks are the alpha rays emitted by the progeny of $^{226}$Ra, i.e.
$^{226}$Ra ($E_{\alpha}=4.784$ MeV), 
$^{222}$Ra ($E_{\alpha}=5.489$ MeV) and $^{218}$Po ($E_{\alpha}=6.002$ MeV).
A small peak at 7.7 MeV is the one emitted by $^{214}$Po ($E_{\alpha}=7.687$ MeV).
The intensity of $^{214}$Po is smaller than the other peaks because its half life is so short as 
164 $\mu$sec that the data acquisition system cannot take all the event.

The yields of each peak were calculated by peak fitting with a Gaussian and linear  function.
From the result of the fitting, the concentrations of radioactivity were obtained as,
\[
\begin{array}{ll}
^{226}\textrm{Ra}  & 105\pm17 \ \mu\textrm{Bq/kg} \\
^{222}\textrm{Rn} & 108\pm17 \ \mu\textrm{Bq/kg} \\
^{218}\textrm{Po} & 100\pm14 \ \mu\textrm{Bq/kg}.
\end{array}
\]
The concentrations of three isotopes agrees with the secular equilibrium in radioactivity.
The other isotopes of U chain and Th chain made no distinct peaks because of the poor energy resolution 
and the small contamination.

The radioactivity of all Th chain progeny was assumed to be same because their half lives are sufficiently 
short.
On the other hand, the U chain is terminated in five segments by isotopes with long half lives as shown below.
\begin{enumerate}
\item $^{238}$U$(4.5\times 10^{9}$y;$\alpha)$ $\rightarrow^{234}$Th(24d;$\beta)$ $\rightarrow^{234}$Pa(1.17m,6.7h;$\beta)$ $\rightarrow^{234}$U$(2.455\times 10^{5}$y;$\alpha$)
\item $^{234}$U$(2.455\times 10^{5}$y;$\alpha)$ $\rightarrow^{230}$Th$(7.538\times 10^{4}$y;
$\alpha$)
\item $^{230}$Th$(7.538\times 10^{4}$y;$\alpha$) $\rightarrow^{226}$Ra(1600y; $\alpha$)
\item $^{226}$Ra(1600y; $\alpha$) $\rightarrow^{222}$Rn(3.82d; $\alpha$) $\rightarrow^{218}$Po
(3.10m; $\alpha$) $\cdots \rightarrow^{214}$Po(164 $\mu$s; $\alpha$) 
$\rightarrow^{210}$Pb(22.4y; $\beta$)
\item $^{210}$Pb(22.4y; $\beta$) $\rightarrow^{210}$Bi(5.01d; $\beta$) $\rightarrow^{210}$Po
(138d; $\alpha$) $\rightarrow^{206}$Pb(stable)
\end{enumerate}
The radioactivity of the isotopes in each segment were assumed to be same.
The yield of the continuous component is the sum of alpha rays whose energies are close to each other.
The yields of neighboring alpha rays were obtained the events in the energy region set by the 
energies and the energy resolutions of the alpha rays.
The energy regions and the yields are listed in table \ref{tb:yields}.
\begin{table}[ht]
\caption{The intensities of each energy regions.The yield is in unit of $10^{-6}$ /sec/kg.}
\label{tb:yields}
\centering
\begin{tabular}{c|rcl} \hline
Region & Energy range (keV) & Nuclei (Chain) & Yield  \\ \hline
A & $3870-4275 $& $^{232}$Th (Th), $^{238}$U (U)& $79\pm6$\\
B & $4524-4996$& $^{226}$Ra (U), $^{230}$Th (U), $^{234}$U (U) &$197\pm9$\\ 
C & $5227-5762$& $^{210}$Pb (U), $^{222}$Ra (U), $^{223}$Th (Th), $^{224}$Ra (Th)&$192\pm9$\\
D & $5775-6365$& $^{218}$Po (U), $^{212}$Bi (Th), $^{220}$Rn (Th)&$134\pm7$\\
E & Gaussian fit& $^{216}$Po (Th) &$13\pm8$\\
 \hline
\end{tabular}
\end{table}
The radioactivity of $^{216}$Po was easily calculated by Gaussian fit.
The radioactivity of the progeny of U chain are derived by subtracting the yield of Th chain intensity.
The result of calculation is listed below as,
\begin{subequations}
\begin{align}
\frac{dN_{^{216}\textrm{Po}}}{dt} & =  13\pm8 \textrm{\ }\mu\textrm{Bq/kg}\\
\frac{dN_{Th-chain}}{dt}  &=\frac{dN_{^{216}\textrm{Po}}}{dt}=13\pm8 \textrm{\ }\mu\textrm{Bq/kg}\\
\frac{dN_{^{218}{\rm Po}}}{dt} & = 108\pm18 \textrm{\ }\mu\textrm{Bq/kg}\\
\frac{dN_{^{226}{\rm Ra}}}{dt}&=\frac{dN_{^{222}{\rm Rn}}}{dt}=108\pm18 \textrm{\ }\mu\textrm{Bq/kg}\\
\frac{dN_{^{210}{\rm Po}}}{dt} & =58\pm26  \textrm{\ }\mu\textrm{Bq/kg}\\
\frac{dN_{^{238}{\rm U}}}{dt} & = 66\pm10  \textrm{\ }\mu\textrm{Bq/kg}\\
\frac{dN_{^{230}{\rm Th}}}{dt}+\frac{dN_{^{234}{\rm U}}}{dt} & = 89\pm21 \textrm{\ }\mu\textrm{Bq/kg}.
\end{align}
\end{subequations}
Where, the intensity of $^{218}$Po was consistent with the one derived by curve fitting.

\section{Discussion}
In the present work, reduction of U chain and Th chain contamination was successfully performed.
The results of purification of NaI(Tl) by the world leading groups are listed in table \ref{tb:compare}.
\begin{table}[ht]
\caption{The concentration of radioactive isotopes in a NaI(Tl) crystal for WIMPs search.}
\label{tb:compare}
\centering
\begin{tabular}{l|rrr} \hline
     & DAMA/LIBRA & DM-ICE & This work \\ \hline
$^{232}$Th & 0.5-0.7 ppt & 2.5 ppt & $3.3\pm2.0$ ppt \\ 
 $^{238}$U & 0.7-10 ppt & 1.4 ppt & $5.4\pm 0.9$ ppt \\
$^{210}$Pb & 5-30 $\mu$Bq/kg & 1470 $\mu$Bq/kg & $58\pm26$ $\mu$Bq/kg \\ \hline
\end{tabular}
\end{table}
The purity of the NaI(Tl) developed by this work reached the same level as the one of DAMA/LIBRA was 
established.
It should be remarked that the small concentration of $^{210}$Pb which is comparable to 
that of DAMA/LIBRA.\@
The effectiveness of resin to reduce a small density of lead ion was successfully verified. 

\begin{figure}[ht]
\centering
\includegraphics[viewport=0 0 717 442, width=0.7\textwidth]{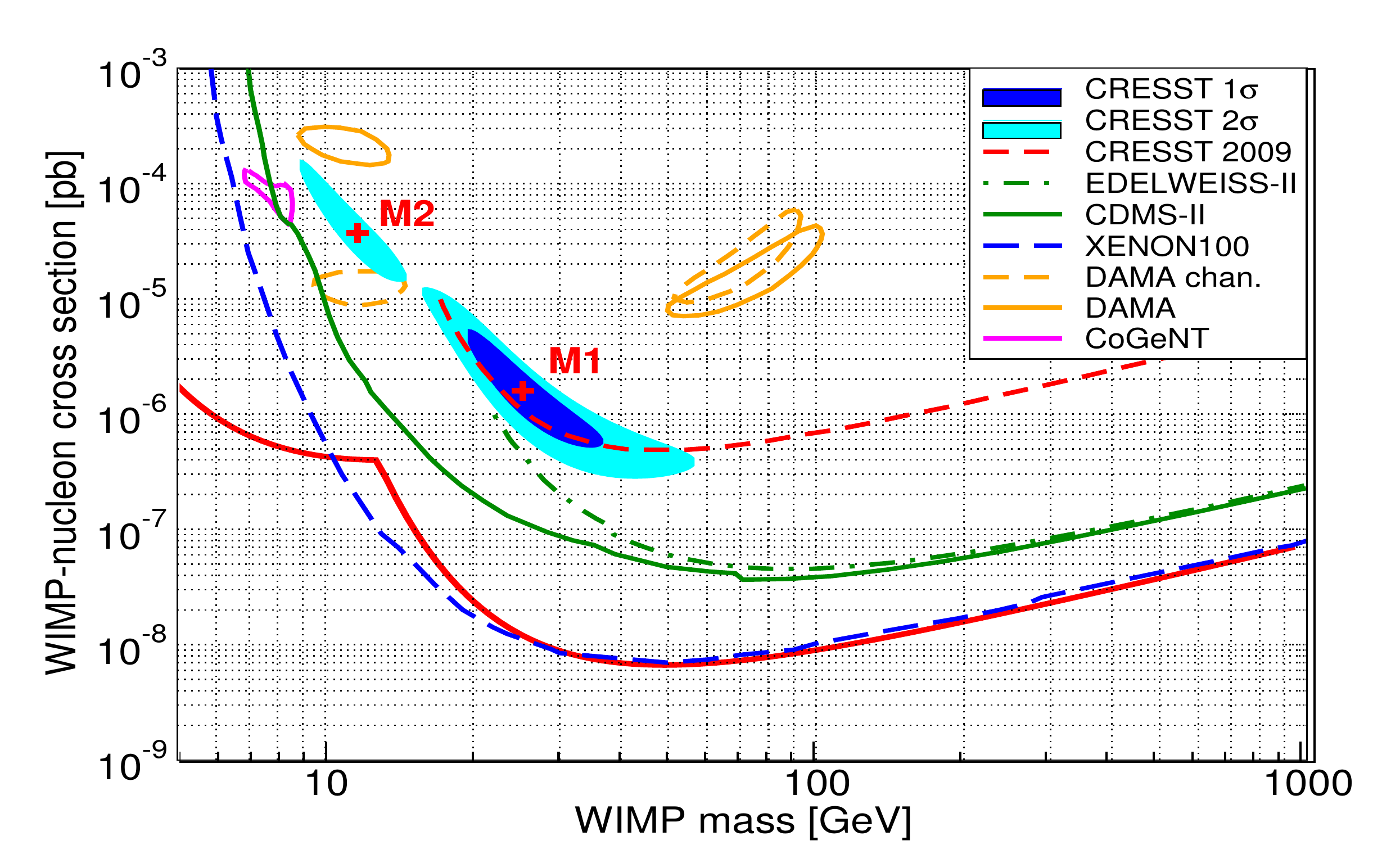}
\caption{The expected sensitivity for spin-independent WIMPs. 
The original sensitivity plot was drawn by CRESST-II\cite{CRESST}
}
\label{fg:furosiki}
\end{figure}

The expected background was simulated by Geant 4.10 \cite{Geant} with the known radioactive impurities in 
the NaI(Tl) crystal and the surrounding materials.
The expected rate at 1 keV in electron equivalent energy was as small as 0.5 /day/keV/kg.
The expected sensitivity to WIMPs candidates is drawn in Fig.\ref{fg:furosiki}.
The red solid line shows the expected sensitivity by 1 ton of our NaI(Tl) for one year.
The KamLAND-PICO project using 1 ton of NaI(Tl) quickly verify not only
the annual modulating signal reported by DAMA/LIBRA
but also the candidates for light WIMPs signal reported by other groups.
\section{Acknowledgment}
The authors thank Professor S.Nakayama for fruitful discussion and encouragement.
This work was supported by Grant-in-Aid for Scientific Research (B) number 24340055.

\end{document}